\documentclass[%
prx,
 amsmath,amssymb,
 aps,
twocolumn,
longbibliography,
showkeys
]{revtex4-1}

\usepackage{graphicx}
\usepackage{epstopdf}
\usepackage{dcolumn}% Align table columns on decimal point
\usepackage{bm}% bold math
\usepackage{color}
\usepackage[mathlines]{lineno}
\usepackage[breaklinks=true,colorlinks=true,linkcolor=blue,urlcolor=blue,citecolor=blue]{hyperref}
\usepackage{ragged2e}
\usepackage{enumitem}
\usepackage{bibunits}
\usepackage{amsthm}
\usepackage{diagbox}

\graphicspath{{figures/}}

\defaultbibliographystyle{elsarticle-num}
\defaultbibliography{biblio}

\renewcommand\thefigure{\arabic{figure}}

\begin{document}

\title[]{Mutation mitigates finite-size effects in spatial evolutionary games}

\author{Chen Shen$^1$}
\email{steven\_shen91@hotmail.com}
\author{Zhixue He$^{2}$}
\author{Lei Shi$^2$}
\email{shi\_lei65@hotmail.com}
\author{Jun Tanimoto$^{1,3}$}
% \email{tanimoto@cm.kyushu-u.ac.jp}

\affiliation{
\vspace{2mm}
\mbox{1. Faculty of Engineering Sciences, Kyushu University, Fukuoka,  816-8580, Japan}
\mbox{2. School of Statistics and Mathematics, Yunnan University of Finance and Economics, Kunming, 650221, China}
\mbox{3. Interdisciplinary Graduate School of Engineering Sciences, Kyushu University, Fukuoka, 816-8580, Japan}
}

\date{\today}

\begin{abstract}
Agent-based simulations are essential for studying cooperation on spatial networks. However, finite-size effects—random fluctuations due to limited network sizes---can cause certain strategies to unexpectedly dominate or disappear, leading to unreliable outcomes. While enlarging network sizes or carefully preparing initial states can reduce these effects, both approaches require significant computational resources. In this study, we demonstrate that incorporating mutation into simulations on limited networks offers an effective and resource-efficient alternative. Using spatial optional public goods games and a more intricate tolerance-based variant, we find that rare mutations preserve inherently stable equilibria. When equilibria are affected by finite-size effects, introducing moderate mutation rates prevent finite-size-induced strategy dominance or extinction, producing results consistent with large-network simulations. Our findings position mutation as a practical tool for improving the reliability of agent-based models and emphasize the importance of mutation sensitivity analysis in managing finite-size effects across spatial networks.

%Agent-based models are extensively applied in evolutionary game theory to study cooperation on spatial networks. However, due to finite-size effects---random fluctuations arising from limited network sizes---stochastic variations in Monte Carlo simulations can lead to unreliable outcomes, as certain strategies may unintentionally dominate or disappear. Increasing network sizes can reduce these stochastic effects by minimizing randomness, but this approach significantly raises computational costs, making large-scale simulations resource-intensive. In this paper, we demonstrate how incorporating simple mutations can effectively counteract these pitfalls without the need for large networks. Using spatial optional public goods games with binary strategy spaces and more complex tolerance-based variants, we find that rare mutations preserve the stability of equilibria in small networks when the equilibrium is genuinely stable. However, when equilibria are affected by finite-size effects, introducing a moderate mutation rate prevents the unintended dominance or loss of strategies, yielding results consistent with large network simulations. Our findings present a cost-effective approach to achieving robust evolutionary outcomes and underscore the importance of mutation sensitivity analysis in managing finite-size effects and enhancing the reliability of agent-based simulations of cooperation on spatial networks.
\end{abstract}

\maketitle

\section*{Introduction}
Understanding how cooperation emerges and persists among unrelated individuals is a fundamental question, as cooperation involves personal costs to benefit others, potentially disadvantaging cooperators relative to selfish individuals~\cite{sachs2004evolution}. This puzzle has been examined across multiple disciplines, each offering unique insights into the survival of cooperation. Experimental economists often study cooperation in one-shot, anonymous scenarios to reveal intrinsic cooperative tendencies. Their theories draw on social norms~\cite{fehr2004social,fehr2018normative}, prosocial preferences~\cite{fehr1999theory,gintis2003explaining}, or even attributions of cooperative behavior to confusion or error~\cite{houser2002revisiting,burton2021payoff}. In contrast, evolutionary game theory investigates how cooperation can emerge and stabilize as an evolutionary equilibrium through mechanisms such as kin selection \cite{eberhard1975evolution}, direct reciprocity \cite{trivers1971evolution}, indirect reciprocity based on reputation \cite{nowak2005evolution}, network reciprocity \cite{nowak1992evolutionary, wang2013insight}, and group selection \cite{henrich2004cultural}. Of these, network reciprocity has been especially influential in explaining how structured populations can sustain cooperation.

Network reciprocity, originally derived from studies on regular lattices, suggests that cooperators can form clusters within networks, supporting each other and resisting exploitation by defectors. This concept has garnered significant attention from researchers in statistical physics, applied mathematics, and computer science, who have further developed theories of cooperation from multiple perspectives. First, they explore how various network structures---such as regular lattices~\cite{hauert2005game}, scale-free networks~\cite{santos2008social}, temporal networks~\cite{li2020evolution}, and higher-order networks~\cite{civilini2024explosive}---affect the conditions under which cooperation survives. A central focus is determining, through mathematical analyses, the conditions under which cooperation persists, often characterized by relationships between the benefit-to-cost ratio and network features like average degree~\cite{ohtsuki2006simple} or degree distribution heterogeneity~\cite{allen2017evolutionary}. Second, they examine whether and how network reciprocity addresses evolutionary challenges that are difficult to resolve in well-mixed populations, such as second-order free-rider problems~\cite{szolnoki2017second} and strong reciprocity, which investigates whether punishing defectors and rewarding cooperators are linked traits in individuals~\cite{szolnoki2013correlation}. Third, they investigate coevolutionary games, where strategies, individual traits, or network structures evolve simultaneously, shedding light on how cooperation persists in complex scenarios~\cite{perc2010coevolutionary}. While analytical methods have advanced in modeling cooperation on spatial networks and identifying survival conditions, they remain constrained by simplified assumptions, such as weak selection. Under weak selection, theorists approximate the natural selection process as a first-order perturbation of neutral drift, essentially ignoring higher-order nonlinear effects on evolution~\cite{ohtsuki2006simple,allen2017evolutionary}. Agent-based models, by contrast, allow for the simulation of complex and nonlinear dynamics without the restrictive assumptions of weak selection, making them essential for studying spatial evolutionary games under a wider range of conditions and providing insights that complement analytical approaches~\cite{adami2016evolutionary}.

%However, a significant challenge in using agent-based models to study phase transitions in spatial evolutionary games is finite-size effects, which can cause deviations in observed critical points and inaccuracies in determining phase transitions.

However, a significant challenge in using agent-based models to study spatial evolutionary games is the finite-size effect. This refers to random fluctuations caused by limited network sizes, which can lead to unexpected strategy extinctions and make results highly sensitive to network size~\cite{szabo2002phase,perc2017statistical,szolnoki2011competition}. Traditional methods to address these effects include carefully preparing initial states~\cite{szolnoki2017second} or conducting comprehensive stability analyses of subsystem competition~\cite{perc2017stability}. While effective, these approaches often require large systems to achieve accurate, size-independent results, significantly increasing computational demands.

In agent-based simulations, mutations are often excluded to observe clear phase transitions, a critical focus in statistical physics for understanding critical phenomena. While this approach simplifies the analysis of dynamic transitions, it poses challenges for evolutionary game theory, where truly stable outcomes must resist invasion by mutant strategies to demonstrate evolutionary stability. Furthermore, relying on large network sizes to mitigate finite-size effects increases computational demands and may not accurately represent real-world systems, which are finite and subject to mutations or behavioral noise.

While mutation has been incorporated in several studies to address finite-size effects in spatial evolutionary games~\cite{lee2024suppressing,li2024antisocial}, the underlying mechanisms remain poorly understood. Specifically, it remains unclear how and why mutation mitigates these effects, and how to determine mutation rates in advance to address finite-size effects for given network sizes. In this study, we systematically explore whether incorporating mutation can mitigate finite-size effects in agent-based simulations with limited network sizes. We begin with a spatial optional public goods game~\cite{szabo2002phase}, where players choose to cooperate by contributing to a common pool, defect by contributing nothing, or act as loners who opt out and receive a fixed payoff. Contributions are multiplied by an enhancement factor and shared among cooperators and defectors. To test the robustness of mutation in addressing finite-size effects, we introduce a more complex tolerance-based variant of the optional public goods game~\cite{szolnoki2016competition,perc2017stability}. In this version, players adopt additional strategies: they cooperate only if the number of defectors in their group remains below a specified tolerance threshold; otherwise, they act as loners. This expansion increases the strategy space to eight distinct strategies, adding complexity to the dynamics.

Focusing on limited network sizes, we first confirm that finite-size effects significantly impact phase boundaries and find that these effects can be effectively mitigated by introducing mutation in agent-based simulations. Specifically, rare mutations preserve equilibrium stability when the equilibria are inherently stable. However, when equilibria are vulnerable to finite-size effects, introducing a moderate mutation rate helps maintain competition among strategies by reintroducing strategies that have occasionally gone extinct, ultimately yielding results comparable to those in large networks. While the efficacy of this approach depends on network size and may not universally resolve finite-size issues, our findings suggest that incorporating mutation offers a cost-effective means to mitigate finite-size effects and achieve reliable results without relying on large networks. Given the difficulty of determining an appropriate network size to counter finite-size effects in advance, especially in the absence of prior analytical results, we propose mutation sensitivity analysis as a general strategy for managing finite-size effects and enhancing the reliability of agent-based simulations.

\section*{Model and Methods}
The agent-based modeling approach for spatial evolutionary games typically comprises four key components: (i) population structure, (ii) game model, (iii) action selection, and (iv) simulation settings. Below, we briefly describe each of these components.

\subsection*{Population structure}
In the simulation, a population of size $N$ is considered and placed on a regular two-dimensional lattice. Each player's neighborhood followed the von Neumann configuration, meaning each player was connected to four neighbors: left, right, up, and down. Periodic boundary conditions are applied, ensuring that players in the last row (or column) are connected to those in the first row (or column) in a lattice.

\subsection*{Game model}
\paragraph{Spatial optional public goods game.}
In the optional public goods game (PGG), cooperators ($C$) contribute an amount $c$ to a common pool, defectors ($D$) contribute nothing, and loners ($L$) opt out of the game, abstaining from the benefits of public goods while receiving a non-negative payoff $\sigma$. The total contributions in the common pool are multiplied by a synergy factor $r$ and distributed equally among participants (excluding loners), regardless of their contributions. The payoff for a player $x$ with strategy $s_x \in \{C, D, L\}$ is given by:
\begin{equation}
\Pi_{s_x} = 
\begin{cases} 
\frac{r c n_c}{n_c + n_d} - c & \text{if } s_x= C \\
\frac{r c n_c}{n_c + n_d} & \text{if } s_x = D \\
\sigma & \text{if } s_x = L 
\end{cases}.
\end{equation}
where $n_c$, $n_d$, and $n_l$ denote the number of cooperators, defectors, and loners within the group, respectively, with $n_c + n_d + n_l = n$, where $n=5$ is the group size. 

In a strict spatial optional PGG, a focal player's payoff  is accumulated over $n = 5$ games, with PGG groups centered around the focal player and each of her four neighbors. In the original spatial optional PGG~\cite{szabo2002phase}, only a simplified scenario is considered, where the payoff is determined by a a single, representative PGG involving the focal player and her four nearest neighbors. Furthermore, when there is only one participating player in the group (i.e., when $n_c+n_d=1$), the player receives the exit payoff $\sigma$, similar to the loners. We adopt these consistent settings of ref.~\cite{szabo2002phase} in our investigation.

% For consistency with the original spatial optional PGG~\cite{szabo2002phase}, we focus solely on a simplified scenario in which the payoff is determined by a single, representative PGG involving the player and their four nearest neighbors. 

\paragraph{tolerance-based variant of optional public goods game.} 
The tolerance-based variant of the optional public goods game introduces eight strategies~\cite{szolnoki2016competition}. In addition to traditional cooperators, defectors, and loners, it includes tolerance-based conditional cooperators ($M_n$). A player adopting strategy $M_n$ acts as a cooperator contribute to common pool if the number of defectors in the group is below their tolerance level 
$n \in \{0, 1, 2, \dots, n-1\}$; otherwise, they act as a loner. Tolerance-based strategies incur a fixed cost $\gamma$, representing the cost of monitoring group composition. 

The total number of contributors ($TC$) is given by:
\begin{equation}
  n_{TC}= n_C + \sum_{n=0}^{n-1} \delta_i n_{M_n}.  
\end{equation}
where $\delta_n=1$ if $n_d$ (the number of defectors) is less than or equal to the tolerance level $n$, and $\delta_n=1$ otherwise. Specifically, a lonely contributor cannot generate the synergistic benefits of collective efforts, so the public pool is not multiplied (equivalent to $r=1$), and the public pool only increases by a factor of $r$ (where 
$1<1<n$) when there are at least two contributors~\cite{szolnoki2016competition}.
The payoff for each strategy in a single PGG is calculated as:
\begin{equation}
    \begin{cases}
P_D = \frac{r ( c \cdot n_{TC}  )}{n_D + n_{TC}} \\
P_C = \Pi_D - c \\
P_L = \sigma \\
P_{M_n} = \delta_n \Pi_C + (1 - \delta_n)\sigma - \gamma
\end{cases}.
\end{equation}
Following Ref.~\cite{szolnoki2016competition}, we adopt strict spatial games. A player’s total payoff $\Pi_{s_x}$ with a strategy $s_x \in \{C, D, L, M_0, M_1, M_2, M_3, M_4\}$ is the sum of payoffs accumulated from $5$ groups, including the player's own group and the groups formed by their four neighbors.

\subsection*{Action selection}
Following Refs.~\cite{szabo2002phase, szolnoki2016competition}, we adopt the imitation updating rule for the spatial optional public goods game and the teaching updating rule for the tolerance-based variant. Both rules are similar but differ in perspective: in the imitation rule, focal players learn the strategy of their neighbors, whereas in the teaching rule, focal players enforce their strategy on their neighbors.

Specifically, let $\Pi_{s_x}$ denote the total payoff of a focal player $x$ and $\Pi_{s_y}$ the payoff of a randomly selected neighbor $y$. The probability of player $x$ imitating the strategy of neighbor $y$ follows the Fermi rule:
\begin{equation}
\centering
w(s_y \rightarrow s_x) = {\frac{1}{1+{\rm exp}[(\Pi_{s_x}-\Pi_{s_y}+\tau)/K]}}.
\label{eq4}
\end{equation}
Similarly, the probability of player $x$ enforcing their strategy $s_x$ on neighbor $y$ (or equivalently, the probability of player $y$ imitating player $x$) is:
\begin{equation}
\centering
w(s_x \rightarrow s_y) = {\frac{1}{1+{\rm exp}[(\Pi_{s_y}-\Pi_{s_x}+\tau)/K]}}.
\label{eq5}
\end{equation}
Here, $\kappa$ represents the irrationality of selection, and $\tau$ is the cost of strategy change. Consistent with prior studies, we set $K=\tau=0.1$ for the spatial optional public goods game, following Ref.~\cite{szabo2002phase}, and $K=0.5, \tau=0$ for the tolerance-based variant, following Ref.~\cite{szolnoki2016competition}.

\paragraph*{Inclusion of mutation.} 
We incorporated mutation, or behavioral noise, into the action selection process. With probability $\mu$, the focal player $x$ mutates to another strategy randomly chosen from the available strategy space. Otherwise, they update their strategy based on Eq.~\ref{eq4} or Eq.~\ref{eq5}, depending on the model.

\subsection*{Simulation settings}
Simulations were conducted over a series of Monte Carlo time steps. In each time step, a focal player was randomly selected to play the game with all their neighbors. Subsequently, one of the focal player’s neighbors was randomly chosen to play the game with their own neighbors. The payoffs of the focal player and the selected neighbor were then compared to determine the outcome: either the focal player imitates the neighbor’s strategy (imitation rule in the optional PGG model) or enforces their own strategy on the neighbor (teaching rule in the tolerance-based variant). Consistent with the setting of previous Refs. ~\cite{szabo2002phase, szolnoki2016competition}, we fixed $c=1$ and $\sigma=1$.

Single simulations ran over $10^5$ time steps, with strategy abundances averaged over the final $5000$ time steps. The population size in the study  typically consisted of $200^2$ players, but varied from $50^2$ to $800^2$ players to to examine the influence of network size. Each data point are averaged from $100$ independent realizations. Throughout this work, we refer to networks with $N=200^2$ or smaller as `small networks', and networks with $N=2000^2$ or larger as `large networks'. Intermediate sizes (e.g., $N=800^2$) are explicitly noted when discussed.

\begin{figure}[t]
    \centering    \includegraphics[width=0.99\linewidth]{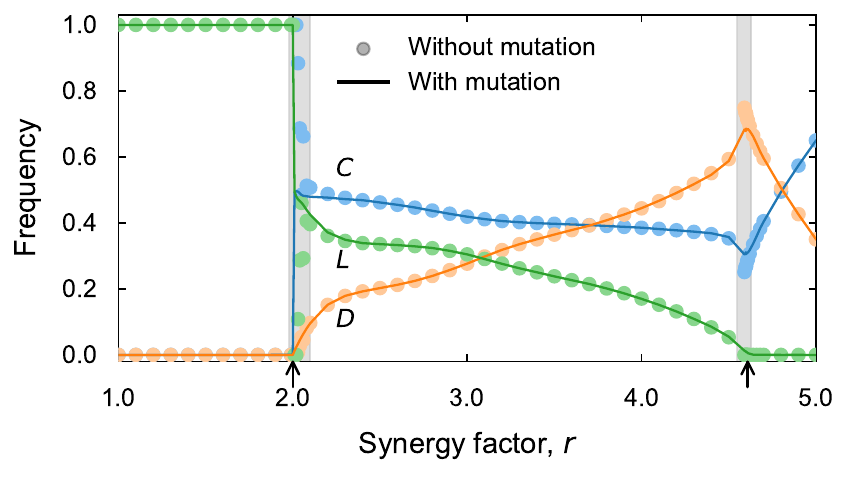}
    \caption{\textbf{In spatial optional public goods games, finite-size effects in small networks near phase transition boundaries are counteracted by introducing mutation, producing results consistent with larger networks.} The figure shows strategy frequencies versus synergy factor $r$ for a small network size $N = 200^2$; dots indicate outcomes without mutation, and solid lines represent outcomes with a mutation rate of $\mu=10^{-6}$. Black arrows at $r=2.0$ and $r=4.6005$ mark critical values from large networks ($N = 2000^2$) as reported in Ref.~\cite{szabo2002phase}, separating phases of pure $L$, $C+D+L$, and $C+D$. Gray regions indicate discrepancies due to finite-size effects.}
    \label{fig_cdl}
\end{figure}

\begin{figure*}[t!]
    \centering    \includegraphics[width=0.91\linewidth]{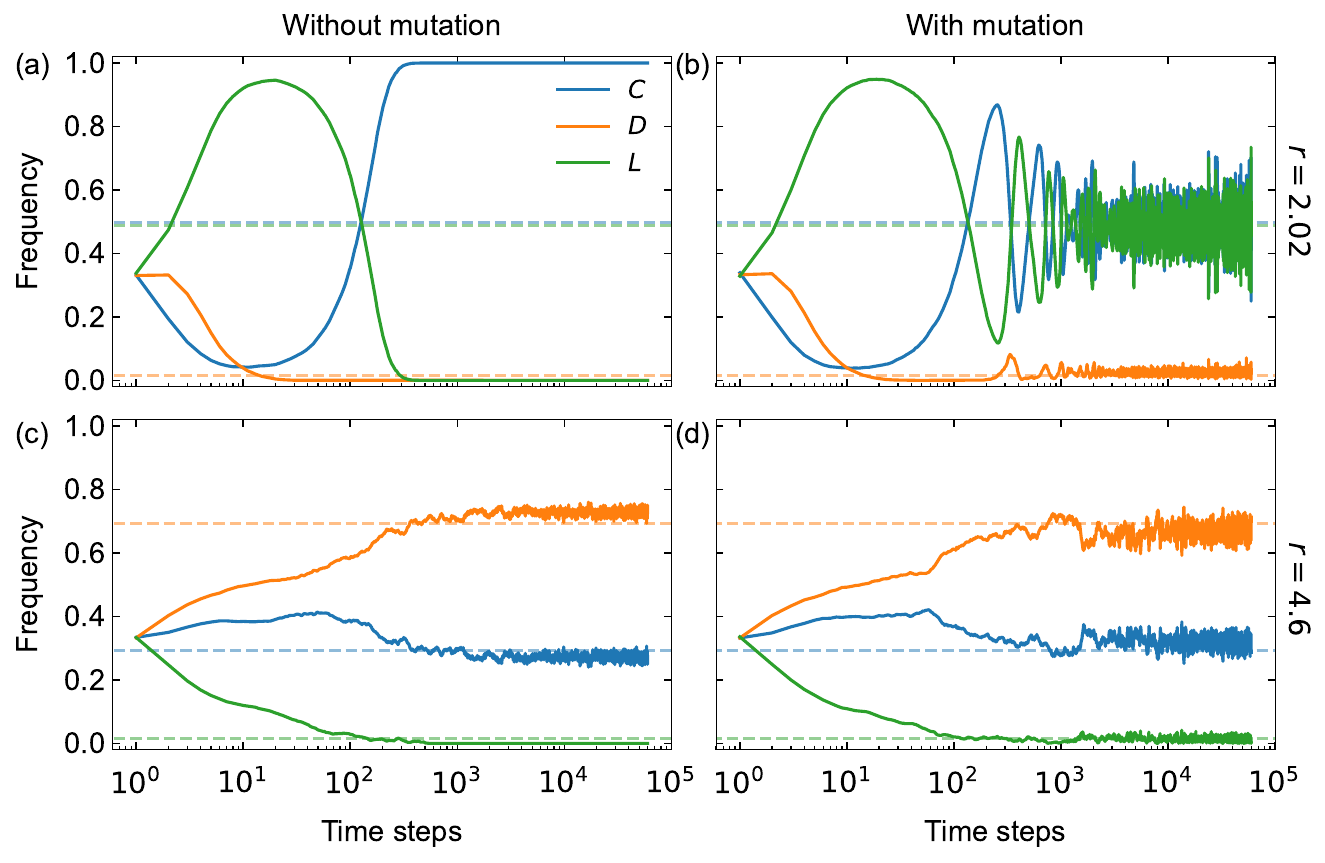}
    \caption{\textbf{In the optional public goods game, including mutation in agent-based simulations restores previously extinct strategies and aligns strategy frequencies with reference results from a network size of $N=2000^2$.} The figure depicts the frequency of each strategy over time for $N=200^2$. Panels (a) and (c) show results without mutation, while (b) and (d) show results with mutation. Top panels use $r=2.02$; bottom panels use $r=4.6$. Parameters are set as $\sigma=1$ and $\mu=10^{-6}$. Horizontal dashed lines denote stable frequencies from large populations ($N=2000^2$) as reported in Ref.~\cite{szabo2002phase}.
    }
    \label{figm2_evo}
\end{figure*}

\begin{figure*}[!t]
    \centering    \includegraphics[width=0.91\linewidth]{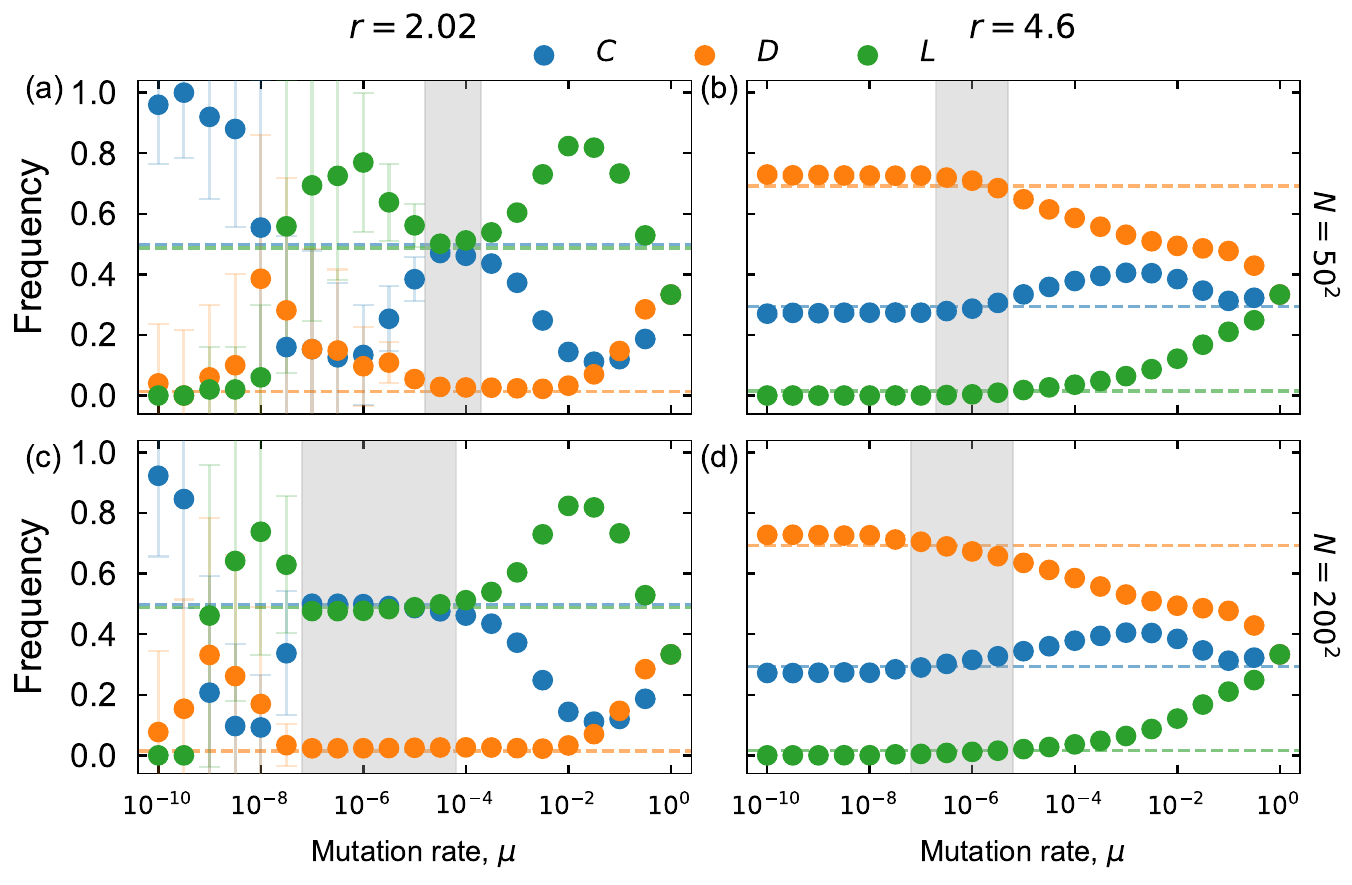}
    \caption{\textbf{Intermediate mutation levels yield strategy frequencies consistent with reference results from large lattices ($N=2000^2$), and the range of effective mutation rates expands with increasing linear system size.} Strategy frequencies are shown as functions of the mutation rate $\mu$ for two network sizes: $N=50^2$ in panels (a) and (b), and $N=200^2$ in panels (c) and (d). Error bars indicate standard deviations across 100 independent runs. The horizontal dashed line represents reference outcomes from a large population ($N=2000^2$), as reported in Ref.~\cite{szabo2002phase}. The shaded region denotes mutation rates where results converge to large population outcomes, effectively mitigating finite-size effects. Parameters are set to $r = 2.02$ for the left panels, and $r = 4.6$ for the right panels.}
    \label{fig_cdl_mu}
\end{figure*}

\begin{figure*}[t!]
    \centering    \includegraphics[width=0.91\linewidth]{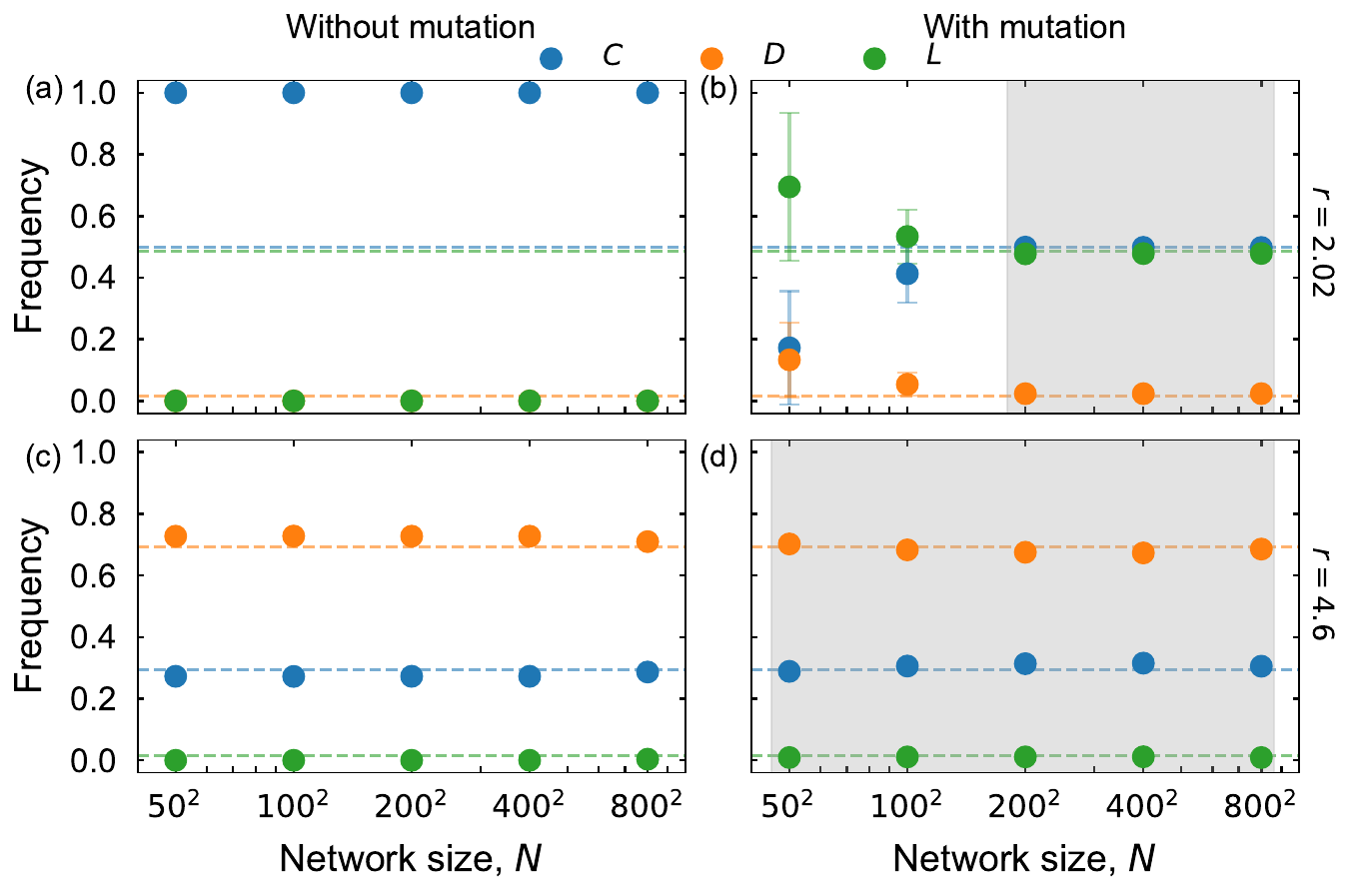}
    \caption{\textbf{In the spatial optional public goods games, the effectiveness of mutation in mitigating finite-size effects remains robust for network sizes above a certain threshold at a fixed mutation rate.} 
    The figure shows the frequencies of each strategy as a function of population size $N$. Error bars represent the standard deviation from 100 independent runs, and the horizontal dashed line shows reference outcomes from a large population ($N=2000^2$) as reported in Ref.~\cite{szabo2002phase}. The shaded region indicates the range of network sizes where mutation results align with those in large populations, effectively resolving finite-size effects. Mutation rate is fixed at $\mu=10^{-6}$. Parameters are $\gamma=2.02$ for panels (a) and (c), and $\gamma=4.6$ for panels (b) and (d).}
    \label{figm2_N}
\end{figure*}

\section*{Results}

\subsection*{Spatial optional public goods game}

Agent-based simulations of the spatial optional public goods game on a $2000 \times 2000$ square lattice without mutation, as conducted in Ref.~\cite{szabo2002phase}, identify precise critical values of the synergy factor $r$ that delineate distinct phases. Specifically, $r = 2$ marks the transition from the pure loner phase to a mixed phase of cooperators, defectors, and loners, while $r = 4.6005$ signifies the shift from the mixed phase to the cooperation-defection phase. These critical values, indicated by arrows in Fig.~\ref{fig_cdl}, serve as benchmark outcomes for comparison.

Building on these findings, we conducted simulations on a smaller $200 \times 200$ lattice to explore the impact of finite-size effects. Without mutation, finite-size effects lead to deviations in critical values, as highlighted by the gray regions in Fig.~\ref{fig_cdl}. Introducing mutation into these smaller lattice simulations effectively mitigates finite-size effects, producing results comparable to those from larger networks. For example, a mutation rate of $\mu = 10^{-6}$ aligns the critical values of $r$ with those observed in the $2000 \times 2000$ lattice. Furthermore, mutation preserves strategy frequencies in regions unaffected by finite-size effects while correcting distortions in the critical regions.

Mutation’s ability to promote strategy diversity is central to its effectiveness. By enabling the reemergence of strategies that occasionally disappear due to finite-size effects, mutation restores evolutionary dynamics and stabilizes strategy levels to match those observed in large networks. At $r=2.02$, finite-size effects in small networks lead to the extinction of defectors (Fig.~\ref{figm2_evo} (a)), while at $r=4.6$, loners may disappear entirely (Fig.~\ref{figm2_evo} (c)). These extinctions disrupt the stable coexistence of three strategies observed in large networks. However, mutation reintroduces these lost strategies, restoring balance. For instance, at $r=2.02$, defectors reintroduced by mutation prevent the dominance of cooperation (Fig.~\ref{figm2_evo} (b)). Similarly, at $r=4.6$, reintroduced loners correct the defector-cooperator coexistence observed in small networks (Fig.~\ref{figm2_evo} (d)). This stabilization ensures that strategy levels closely align with those observed in large networks.

A detailed analysis of varying mutation levels reveals that certain mutation rates effectively mitigate finite-size effects, with the range of effective mutation rates expanding as the linear system size increases. As shown in Fig.~\ref{fig_cdl_mu}, a `window of mutations' exists where the results closely match the reference outcomes from a $2000 \times 2000$ lattice. For an $N=50^2$ lattice, effective mutation rates range from approximately $10^{-5}$ to $10^{-3}$ at $r=2.02$ and from about $5 \times 10^{-6}$ to $10^{-5}$ at $r=4.6$. For a larger $N=200^2$ lattice, these ranges shift to approximately $10^{-7}$ to $10^{-4}$ at $r=2.02$ and remain roughly $10^{-7}$ to $10^{-5}$ at $r=4.6$. The upper limit indicates where mutations begin to significantly alter the system's dynamics, while the lower threshold represents the minimum mutation rate required to address finite-size effects. Mutation rates outside these ranges result in strategy frequencies that deviate significantly from those observed in large networks. Comparisons between $N=50^2$ and $N=200^2$ reveal that the valid mutation window broadens as the system size increases. This finding suggests that beyond a certain system size, the mutation rate required to mitigate finite-size effects stabilizes at a fixed range, allowing reliable results without further adjustments.

Incorporating fixed intermediate mutation rates significantly reduces the network size required to mitigate finite-size effects, though the effectiveness of these rates varies with network size. Without mutation, finite-size effects persist at $r=2.02$ even as lattice size increases from $N=50^2$ to $N=800^2$. While results at $r=2.02$ remain stable for $N \leq 800^2$ (Fig.~\ref{figm2_N} (a)), they still deviate from those observed on a $2000 \times 2000$ lattice. At $r=4.6$, finite-size effects diminish at $N=800^2$, where results align closely with those from $2000 \times 2000$, but small discrepancies persist for $N < 800^2$ (Fig.~\ref{figm2_N} (c)). These findings highlight the difficulty of mitigating finite-size effects solely by increasing network size, as determining an appropriate lattice size can be challenging, and residual discrepancies may persist even when results appear stable. In contrast, incorporating mutation achieves consistent results with much smaller networks, requiring only $N=200^2$ at $r=2.02$ and $N=50^2$ at $r=4.6$ (Fig.~\ref{figm2_N}(b) and (d)). Moreover, with mutation, results remain stable for network sizes above approximately $N=200^2$.

Overall, these results demonstrate that incorporating mutation in agent-based simulations with small network sizes effectively mitigates finite-size effects, producing outcomes consistent with the referenced results from large networks (e.g., $N=2000^2$) adapted from Ref.~\cite{szabo2002phase}. However, mutation is not a universal solution; its effectiveness depends on factors such as network size, mutation rate, and key game parameters like the synergy factor. To further validate the robustness and generality of these findings, we extend this approach to the spatial tolerance-based variant of optional public goods games in the subsequent analysis.

\subsection*{Robustness check: tolerance-based variant of optional public goods game}

The tolerance-based variant of public goods games (PGGs), with its eight competing strategies, introduces significantly more complex evolutionary dynamics and phase transitions compared to the simpler three-strategy optional PGG. Simulations reported in Ref.~\cite{szolnoki2016competition} used lattice sizes of $2400 \times 2400$ or larger to ensure that results were independent of network size, reflecting the need for large-scale simulations in the absence of mutation.

Despite the added complexity, our comparative analysis demonstrates that the results of the tolerance-based variant closely align with those of the simpler spatial optional PGG model (Figures~\ref{figm1_cross}-\ref{figm1_n} in the Supplementary Information), with two notable exceptions. First, incorporating a mutation rate of $\mu=10^{-6}$ in agent-based simulations on a square lattice of size $200^2$ does not fully resolve finite-size effects, leaving a gap compared to results obtained from a $2400^2$ lattice. To conserve computational resources, we did not extend simulations to match the outcomes from the larger lattice. However, adjusting both the network size and mutation rate is expected to achieve consistency with the referenced results, as suggested by Figures~\ref{figm1_mu} and \ref{figm1_n} in the Supplementary Information. Second, for smaller network sizes (e.g., $N=50^2$, as shown in Figure~\ref{figm1_mu}(b)), identifying an intermediate mutation range that reproduces results consistent with larger networks is not feasible. These observations highlight the critical importance of carefully selecting network size and mutation rates when using mutation to address finite-size effects.

Nevertheless, the tolerance-based variant reinforces our main conclusion: mutation enables agent-based simulations on small networks to effectively mitigate finite-size effects, offering a robust approach even in systems with complex evolutionary dynamics. In practice, determining network-size-independent results is often challenging, particularly in the absence of prior analytical insights. To address this, we recommend performing mutation-sensitivity analyses across a range of network sizes in spatial evolutionary games. This approach ensures robust mitigation of finite-size effects without requiring excessively large networks, as in the method of solely increasing network size, and enhances the reliability of simulation outcomes, even in the face of complex dynamics.

\section*{Discussions}
Our study addresses finite-size effects in agent-based modeling of spatial evolutionary games by introducing mutation into simulations with small network sizes. Using the spatial public goods game and its tolerance-based variant as examples, we demonstrate that introducing mutation into small networks offers a simple and cost-effective alternative to resource-intensive large networks~\cite{perc2017stability,szolnoki2017second}. We found that rare mutations do not alter evolutionary outcomes when these are stable. When equilibria are affected by finite-size effects, intermediate mutation levels yield results consistent with those from large networks, effectively counteracting finite-size effects. Notably, the effectiveness of including mutation in agent-based simulations depends on both network size and mutation rate. Given the difficulty of determining network-independent results in advance, especially without prior analytical benchmarks, incorporating mutation and performing a mutation-sensitivity analysis across different network sizes offers a practical and efficient approach to ensure reliable outcomes.

The implications of our findings extend beyond the optional public goods game and its specific variant to a broader class of spatial evolutionary games. We focused on the spatial optional public goods game because it is a foundational and classical model that exemplifies finite-size effects and underscores the necessity of large network sizes to mitigate them. The tolerance-based variant, with its eight competing strategies, further underscores the challenges of finite-size effects and serves as a robust test case for evaluating the solutions of subsystem competition~\cite{perc2017stability}. Finite-size effects are also prevalent in rock-paper-scissors-type Lotka–Volterra systems~\cite{frachebourg1996spatial,frachebourg1998fixation,provata2003spontaneous} and other multi-strategy evolutionary games~\cite{szolnoki2014cyclic,perc2017statistical}, such as public goods games with peer or pool punishment and multi-strategy extensions of the prisoner’s dilemma. Although we did not extend our approach to other spatial evolutionary games, we propose that incorporating mutation to address finite-size effects is broadly applicable and independent of specific game details. By reintroducing extinct strategies and maintaining fair competition, mutation promotes strategy diversity, offering a viable alternative to relying solely on large networks to enhance strategy survival. Future studies could explore the applicability of this approach to other spatial evolutionary game models and further investigate its potential.

Introducing mutation introduces stochasticity that prevents strategies from going extinct, which may obscure exact phase diagrams and phase transitions---a central focus in statistical physics---but facilitates the study of more complex spatial evolutionary games. While concepts such as phase transitions and self-organization provide valuable insights into complex systems, they often restrict spatial evolutionary game studies to small strategy spaces, typically with four or fewer strategies. This limitation arises because determining phase transition points in games with more complex strategy spaces is challenging due to finite-size effects, particularly in the absence of analytical methods. By shifting the focus from strictly physical phenomena and employing mutation-sensitive analysis to enhance the reliability of agent-based simulations, researchers can explore more intricate evolutionary games on structured networks without requiring extensive computational resources to overcome finite-size effects. For instance, in well-mixed populations, the optional public goods game with a comprehensive punishment set—comprising 24 strategies where cooperators, defectors, and loners can punish one another—shows high levels of antisocial punishment, exceeding those observed in games with limited punishment sets~\cite{rand2011evolution}. While such games have been extensively studied in well-mixed populations, their counterparts on structured networks remain poorly understood and are often constrained to simplified punishment strategy spaces~\cite{szolnoki2017second}. We are currently applying this approach to study these complex strategy sets on structured networks, and our work is in progress.

Recent research on direct reciprocity, which traditionally assumes rare mutations, has shown that relaxing this assumption reveals the hidden cooperation-promoting role of mutation~\cite{tkadlec2023mutation}. Together with our findings, this underscores mutation's critical yet often overlooked role in shaping evolutionary dynamics, especially in simplified models where mutation is neglected. Future studies should move beyond mutation-free or rare-mutation assumptions to embrace more realistic scenarios, which may reveal previously hidden outcomes. We hope our results inspire further exploration of complex spatial evolutionary games and deepen our understanding of cooperation in structured populations.

Our findings have broader implications for addressing finite-size effects in fields like network synchronization and disease spreading~\cite{porter2016dynamical}. Traditional methods in these areas include finite-size scaling to understand system-size dependencies and enlarging networks to reduce fluctuations~\cite{privman1990finite,son2010thermal,karrer2011competing,pastor2002epidemic}. Since mutation in evolutionary games mitigates finite-size effects by maintaining strategy diversity, could analogous mechanisms promoting diversity or stochasticity be applied here? For example, might variability in oscillator phases or coupling strengths enhance synchronization in small networks? Could heterogeneity in transmission rates or recovery times influence epidemic thresholds in small populations? Exploring these possibilities may deepen our understanding of finite-size effects across disciplines and lead to more robust models in network dynamics.

%Beyond evolutionary game theory, our findings provide implications for addressing finite-size effects in related fields such as network synchronization~\cite{marodi2002synchronization,son2010thermal} and disease spreading~\cite{karrer2011competing}, where amplified fluctuations in small networks pose significant challenges. Stochastic elements analogous to mutation, such as variability in oscillator phases or noise in transmission rates, could similarly mitigate these effects and align small-network dynamics with those of larger networks. Although the mechanisms differ, with mutation enhancing strategy diversity and noise influencing synchronization stability or epidemic thresholds, the shared principle of using stochasticity to stabilize dynamics suggests broader applicability. This warrants further exploration to extend our understanding of finite-size effects across disciplines.

\paragraph*{Acknowledgments}
We thank Prof. Attila Szolnoki for reviewing the first draft and providing constructive suggestions, particularly for the interpretation of Figure 3. We also thank Dr. Yini Geng and Dr. Zhenyu Shi for useful discussions. We also acknowledge the support provided by (i) the National Natural Science Foundation of China (grant Nos. 12271471 and 11931015), Major Program of National Fund of Philosophy and Social Science of China (grants Nos. 22\&ZD158 and 22VRCO49) to L.\, S.; (ii) JSPS KAKENHI (Grant no. JP 23H03499) to C.\,S.; (iii) Yunnan Provincial Department of Education Science Research Fund Project (Grant No.~2024Y503) to to Z.\,H; and (iv) the grant-in-Aid for Scientific Research from JSPS, Japan, KAKENHI (Grant No. JP 20H02314 and JP 23H03499) awarded to J.\,T.

\paragraph*{Author contributions}
C.S. conceived, designed the study; C.S. and Z.H. performed research; C.S. Z.H. L.S. and J.T. analyzed results and wrote manuscript.

\paragraph*{Competing interest} Authors declare that they have no conflict of interests.

\paragraph*{Data availability}
The code that support the findings of this study is freely accessible at OSF: \url{https://osf.io/nw378/}, Ref. \cite{shen_osf_2024}.

\bibliography{refs}

\begin{thebibliography}{10}

\bibitem{sachs2004evolution}
Joel~L Sachs, Ulrich~G Mueller, Thomas~P Wilcox, and James~J Bull.
\newblock The evolution of cooperation.
\newblock {\em The Quarterly Review of Biology}, 79(2):135--160, 2004.

\bibitem{fehr2004social}
Ernst Fehr and Urs Fischbacher.
\newblock Social norms and human cooperation.
\newblock {\em Trends in Cognitive Sciences}, 8(4):185--190, 2004.

\bibitem{fehr2018normative}
Ernst Fehr and Ivo Schurtenberger.
\newblock Normative foundations of human cooperation.
\newblock {\em Nature Human Behaviour}, 2(7):458--468, 2018.

\bibitem{fehr1999theory}
Ernst Fehr and Klaus~M Schmidt.
\newblock A theory of fairness, competition, and cooperation.
\newblock {\em The Quarterly Journal of Economics}, 114(3):817--868, 1999.

\bibitem{gintis2003explaining}
Herbert Gintis, Samuel Bowles, Robert Boyd, and Ernst Fehr.
\newblock Explaining altruistic behavior in humans.
\newblock {\em Evolution and Human Behavior}, 24(3):153--172, 2003.

\bibitem{houser2002revisiting}
Daniel Houser and Robert Kurzban.
\newblock Revisiting kindness and confusion in public goods experiments.
\newblock {\em American Economic Review}, 92(4):1062--1069, 2002.

\bibitem{burton2021payoff}
Maxwell~N Burton-Chellew and Stuart~A West.
\newblock Payoff-based learning best explains the rate of decline in cooperation across 237 public-goods games.
\newblock {\em Nature Human Behaviour}, 5(10):1330--1338, 2021.

\bibitem{eberhard1975evolution}
Mary Jane~West Eberhard.
\newblock The evolution of social behavior by kin selection.
\newblock {\em The Quarterly Review of Biology}, 50(1):1--33, 1975.

\bibitem{trivers1971evolution}
Robert~L Trivers.
\newblock The evolution of reciprocal altruism.
\newblock {\em The Quarterly Review of Biology}, 46(1):35--57, 1971.

\bibitem{nowak2005evolution}
Martin~A Nowak and Karl Sigmund.
\newblock Evolution of indirect reciprocity.
\newblock {\em Nature}, 437(7063):1291--1298, 2005.

\bibitem{nowak1992evolutionary}
Martin~A Nowak and Robert~M May.
\newblock Evolutionary games and spatial chaos.
\newblock {\em Nature}, 359(6398):826--829, 1992.

\bibitem{wang2013insight}
Zhen Wang, Satoshi Kokubo, Jun Tanimoto, Eriko Fukuda, and Keizo Shigaki.
\newblock Insight into the so-called spatial reciprocity.
\newblock {\em Physical Review E}, 88(4):042145, 2013.

\bibitem{henrich2004cultural}
Joseph Henrich.
\newblock Cultural group selection, coevolutionary processes and large-scale cooperation.
\newblock {\em Journal of Economic Behavior \& Organization}, 53(1):3--35, 2004.

\bibitem{hauert2005game}
Christoph Hauert and Gy{\"o}rgy Szab{\'o}.
\newblock Game theory and physics.
\newblock {\em American Journal of Physics}, 73(5):405--414, 2005.

\bibitem{santos2008social}
Francisco~C Santos, Marta~D Santos, and Jorge~M Pacheco.
\newblock Social diversity promotes the emergence of cooperation in public goods games.
\newblock {\em Nature}, 454(7201):213--216, 2008.

\bibitem{li2020evolution}
Aming Li, Lei Zhou, Qi~Su, Sean~P Cornelius, Yang-Yu Liu, Long Wang, and Simon~A Levin.
\newblock Evolution of cooperation on temporal networks.
\newblock {\em Nature Communications}, 11(1):2259, 2020.

\bibitem{civilini2024explosive}
Andrea Civilini, Onkar Sadekar, Federico Battiston, Jes{\'u}s G{\'o}mez-Garde{\~n}es, and Vito Latora.
\newblock Explosive cooperation in social dilemmas on higher-order networks.
\newblock {\em Physical Review Letters}, 132(16):167401, 2024.

\bibitem{ohtsuki2006simple}
Hisashi Ohtsuki, Christoph Hauert, Erez Lieberman, and Martin~A Nowak.
\newblock A simple rule for the evolution of cooperation on graphs and social networks.
\newblock {\em Nature}, 441(7092):502--505, 2006.

\bibitem{allen2017evolutionary}
Benjamin Allen, Gabor Lippner, Yu-Ting Chen, Babak Fotouhi, Naghmeh Momeni, Shing-Tung Yau, and Martin~A Nowak.
\newblock Evolutionary dynamics on any population structure.
\newblock {\em Nature}, 544(7649):227--230, 2017.

\bibitem{szolnoki2017second}
Attila Szolnoki and Matja{\v{z}} Perc.
\newblock Second-order free-riding on antisocial punishment restores the effectiveness of prosocial punishment.
\newblock {\em Physical Review X}, 7(4):041027, 2017.

\bibitem{szolnoki2013correlation}
Attila Szolnoki and Matja{\v{z}} Perc.
\newblock Correlation of positive and negative reciprocity fails to confer an evolutionary advantage: Phase transitions to elementary strategies.
\newblock {\em Physical Review X}, 3(4):041021, 2013.

\bibitem{perc2010coevolutionary}
Matja{\v{z}} Perc and Attila Szolnoki.
\newblock Coevolutionary games—a mini review.
\newblock {\em BioSystems}, 99(2):109--125, 2010.

\bibitem{adami2016evolutionary}
Christoph Adami, Jory Schossau, and Arend Hintze.
\newblock Evolutionary game theory using agent-based methods.
\newblock {\em Physics of Life Reviews}, 19:1--26, 2016.

\bibitem{szabo2002phase}
Gy{\"o}rgy Szab{\'o} and Christoph Hauert.
\newblock Phase transitions and volunteering in spatial public goods games.
\newblock {\em Physical Review Letters}, 89(11):118101, 2002.

\bibitem{perc2017statistical}
Matja{\v{z}} Perc, Jillian~J Jordan, David~G Rand, Zhen Wang, Stefano Boccaletti, and Attila Szolnoki.
\newblock Statistical physics of human cooperation.
\newblock {\em Physics Reports}, 687:1--51, 2017.

\bibitem{szolnoki2011competition}
Attila Szolnoki, Gy{\"o}rgy Szab{\'o}, and Lilla Czak{\'o}.
\newblock Competition of individual and institutional punishments in spatial public goods games.
\newblock {\em Physical Review E}, 84(4):046106, 2011.

\bibitem{perc2017stability}
Matja{\v{z}} Perc.
\newblock Stability of subsystem solutions in agent-based models.
\newblock {\em European Journal of Physics}, 39(1):014001, 2017.

\bibitem{lee2024suppressing}
Hsuan-Wei Lee, Colin Cleveland, and Attila Szolnoki.
\newblock Suppressing defection by increasing temptation: The impact of smart cooperators on a social dilemma situation.
\newblock {\em Applied Mathematics and Computation}, 479:128864, 2024.

\bibitem{li2024antisocial}
Shulan Li, Chunpeng Du, Xingxu Li, Chen Shen, and Lei Shi.
\newblock Antisocial peer exclusion does not eliminate the effectiveness of prosocial peer exclusion in structured populations.
\newblock {\em Journal of Theoretical Biology}, 576:111665, 2024.

\bibitem{szolnoki2016competition}
Attila Szolnoki and Matja{\v{z}} Perc.
\newblock Competition of tolerant strategies in the spatial public goods game.
\newblock {\em New Journal of Physics}, 18(8):083021, 2016.

\bibitem{frachebourg1996spatial}
Laurent Frachebourg, Paul~L Krapivsky, and Eli Ben-Naim.
\newblock Spatial organization in cyclic lotka-volterra systems.
\newblock {\em Physical Review E}, 54(6):6186, 1996.

\bibitem{frachebourg1998fixation}
L~Frachebourg and PL~Krapivsky.
\newblock Fixation in a cyclic lotka-volterra model.
\newblock {\em Journal of Physics A: Mathematical and General}, 31(15):L287, 1998.

\bibitem{provata2003spontaneous}
A~Provata and GA~Tsekouras.
\newblock Spontaneous formation of dynamical patterns with fractal fronts in the cyclic lattice lotka-volterra model.
\newblock {\em Physical Review E}, 67(5):056602, 2003.

\bibitem{szolnoki2014cyclic}
Attila Szolnoki, Mauro Mobilia, Luo-Luo Jiang, Bartosz Szczesny, Alastair~M Rucklidge, and Matja{\v{z}} Perc.
\newblock Cyclic dominance in evolutionary games: a review.
\newblock {\em Journal of the Royal Society Interface}, 11(100):20140735, 2014.

\bibitem{rand2011evolution}
David~G Rand and Martin~A Nowak.
\newblock The evolution of antisocial punishment in optional public goods games.
\newblock {\em Nature Communications}, 2(1):434, 2011.

\bibitem{tkadlec2023mutation}
Josef Tkadlec, Christian Hilbe, and Martin~A Nowak.
\newblock Mutation enhances cooperation in direct reciprocity.
\newblock {\em Proceedings of the National Academy of Sciences}, 120(20):e2221080120, 2023.

\bibitem{porter2016dynamical}
Mason~A Porter and James~P Gleeson.
\newblock Dynamical systems on networks.
\newblock {\em Frontiers in Applied Dynamical Systems: Reviews and Tutorials}, 4:29, 2016.

\bibitem{privman1990finite}
Vladimir Privman.
\newblock {\em Finite size scaling and numerical simulation of statistical systems}.
\newblock World Scientific, 1990.

\bibitem{son2010thermal}
Seung-Woo Son and Hyunsuk Hong.
\newblock Thermal fluctuation effects on finite-size scaling of synchronization.
\newblock {\em Physical Review E}, 81(6):061125, 2010.

\bibitem{karrer2011competing}
Brian Karrer and Mark~EJ Newman.
\newblock Competing epidemics on complex networks.
\newblock {\em Physical Review E}, 84(3):036106, 2011.

\bibitem{pastor2002epidemic}
Romualdo Pastor-Satorras and Alessandro Vespignani.
\newblock Epidemic dynamics in finite size scale-free networks.
\newblock {\em Physical Review E}, 65(3):035108, 2002.

\bibitem{shen_osf_2024}
Chen Shen, Zhixue He, Lei Shi, and Jun Tanimoto.
\newblock Mutation mitigates finite-size effects in spatial evolutionary games, Nov 2024.

\end{thebibliography}
\bibliographystyle{unsrt}

\clearpage
\onecolumngrid
\setcounter{figure}{0}
\renewcommand\thefigure{S\arabic{figure}}
\setcounter{page}{1}
\renewcommand\thepage{S\arabic{page}}

\section*{
Supplementary Information for\\
``Mutation mitigates finite-size effects in spatial evolutionary games''}

\begin{figure*}[!h]
    \centering    \includegraphics[width=0.95\linewidth]{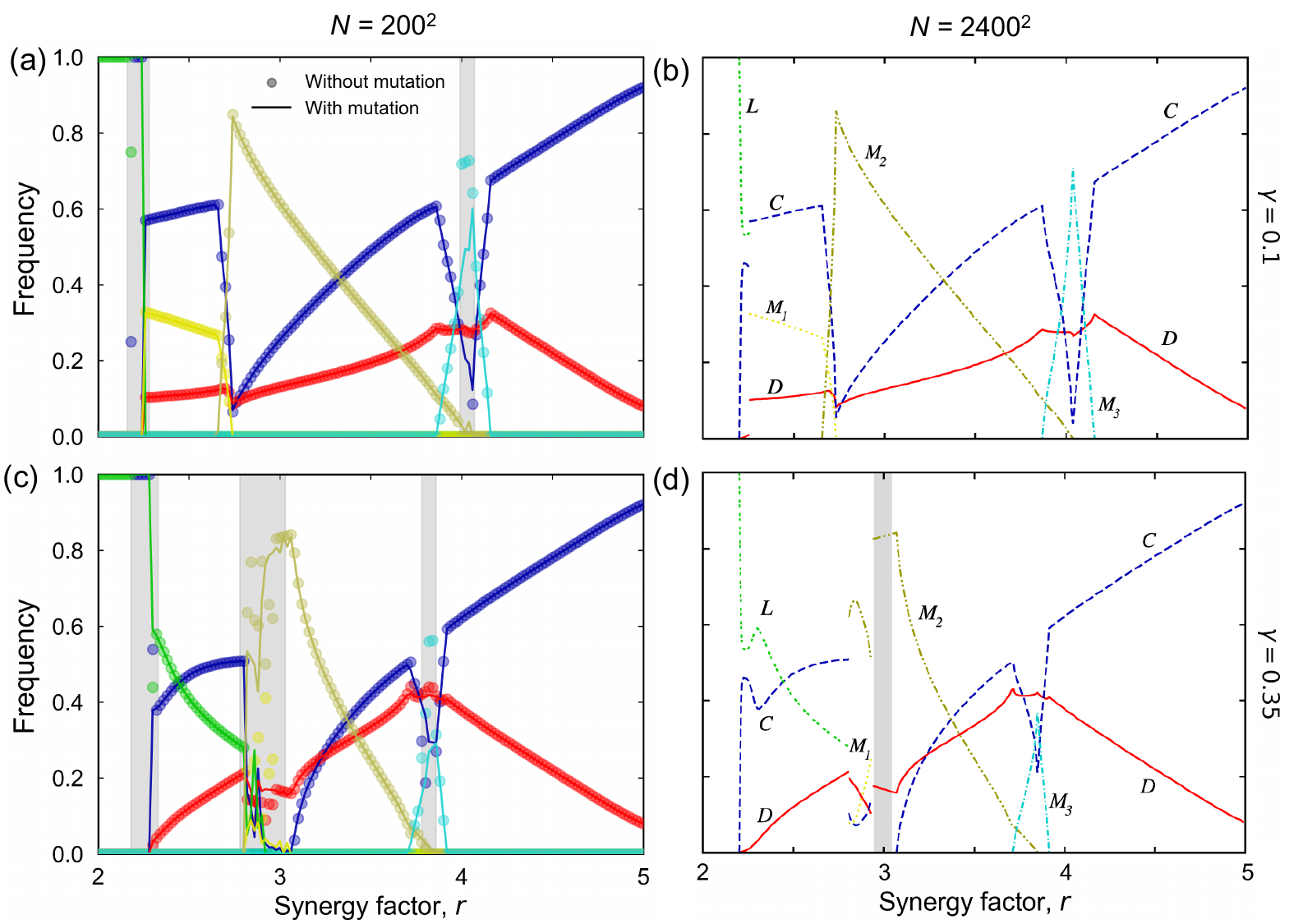}
    \caption{ 
    \textbf{In the tolerance-based variant of PGGs, mutation addresses finite-size effects, particularly along phase boundaries.} The figure shows the frequencies of each strategy as a function of the synergy factor $r$. Panels (a) and (c) present results for a small population size $N = 200^2$; dots indicate outcomes without mutation, and solid lines show outcomes with mutation. Panels (b) and (d) display reference results for a large population size $N = 2400^2$ and are adapted from Ref.~\cite{szolnoki2016competition}. The results with mutation in the small population closely align with the reference results. Gray areas in (a) and (c) highlight finite-size effects, showing discrepancies between small and large network results. Parameters for panels (a) and (b) are set to $\gamma = 0.1$ and $\gamma = 0.35$ for panels (c) and (d). Other parameters are set as $\sigma = 1$, $\kappa=0.5$, $\mu = 10^{-6}$ and $\tau=0$.
    Panels (b) and (d) reprinted from Ref.~\cite{szolnoki2016competition} under the Creative Commons Attribution 3.0 Unported (CC BY 3.0).
}
    \label{figm1_cross}
\end{figure*}

\begin{figure*}[t!]
    \centering    \includegraphics[width=0.95\linewidth]{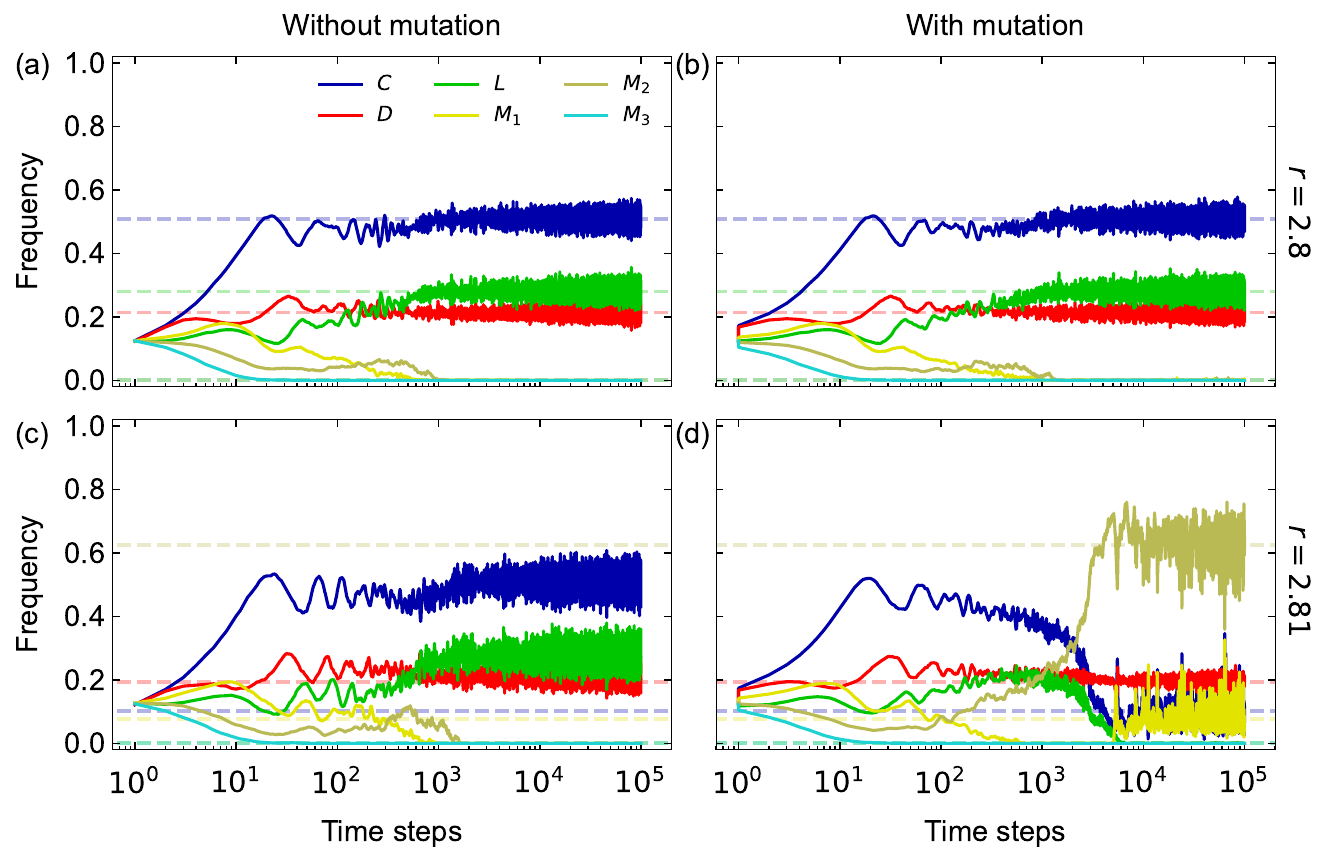}
    \caption{ 
    \textbf{In the tolerance-based variant of PGGs, mutation preserves stable outcomes and revives extinct strategies in small networks, ensuring fair competition and consistency with large-network results.} The figure depicts the frequency of each strategy over time for $N=200^2$. Panels (a) and (c) show results without mutation, while (b) and (d) show results with mutation. Top panels use $r=2.8$; bottom panels use $r=2.81$. Parameters are set as $\sigma=1$, $\gamma=0.35$, $\kappa=0.5$, $\mu=10^{-6}$ and $\tau=0$. Horizontal dashed lines denote stable frequencies from large populations ($N=2400^2$) as reported in Ref.~\cite{szolnoki2016competition}. 
    }
    \label{figm1_evo}
\end{figure*}

\begin{figure*}[t!]
    \centering    \includegraphics[width=0.95\linewidth]{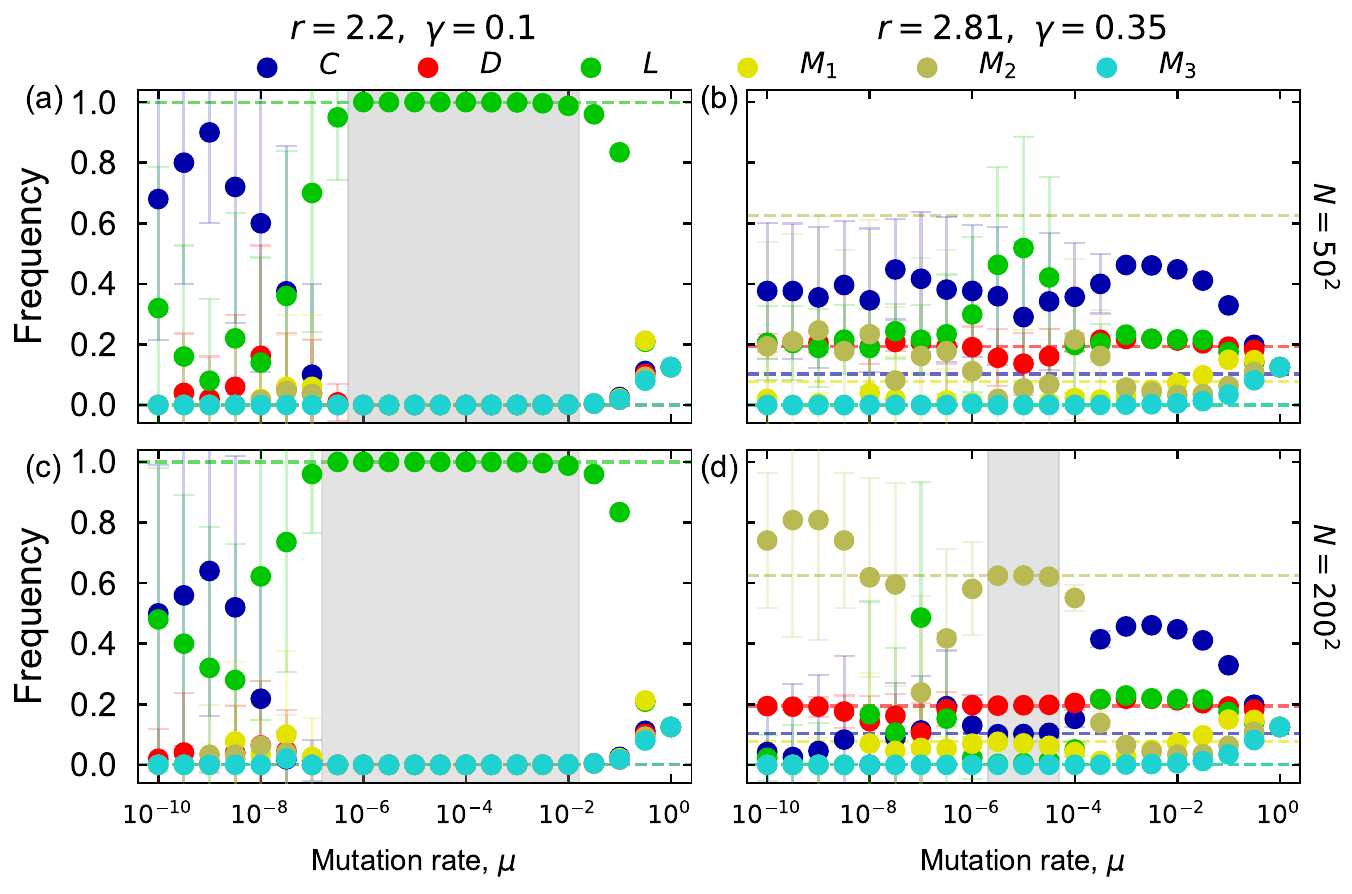}
    \caption{ 
    \textbf{Mutation effectively mitigates finite-size effects in the tolerance-based variant of PGGs.} Mutation at intermediate rates and appropriate network sizes effectively addresses finite-size issues in small networks. Strategy frequencies are shown as functions of the mutation rate $\mu$ for two network sizes: $N=50^2$ in panels (a) and (b), and $N=200^2$ in panels (c) and (d). Error bars indicate standard deviations across 100 independent runs. The horizontal dashed line represents reference outcomes from a large population ($N=2400^2$), as reported in Ref.~\cite{szolnoki2016competition}. The shaded region denotes mutation rates where results converge to large population outcomes, effectively mitigating finite-size effects. Parameters are set as $r = 2.2$, $\gamma = 0.1$ for the left panels, and $r = 2.81$, $\gamma = 0.35$ for the right panels.}
    \label{figm1_mu}
\end{figure*}

\begin{figure*}[t!]
    \centering    \includegraphics[width=0.95\linewidth]{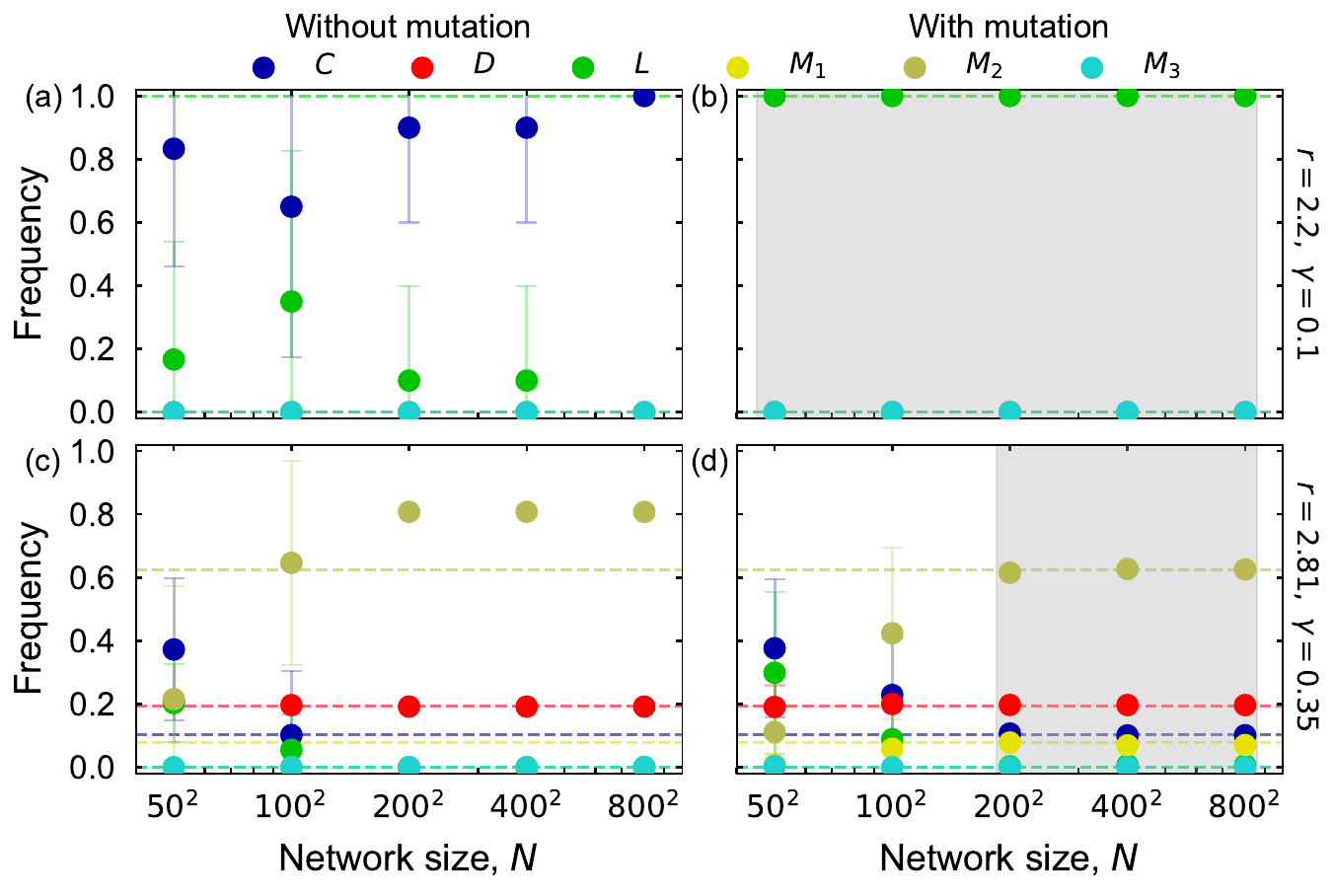}
    \caption{ 
    \textbf{In the tolerance-based variant of PGGs, the effectiveness of mutation in mitigating finite-size effects remains robust for network sizes above a certain threshold at a fixed mutation rate.} The figure shows the frequencies of each strategy as a function of population size $N$. Error bars represent the standard deviation from 100 independent runs, and the horizontal dashed line shows reference outcomes from a large population ($N=2400^2$) as reported in Ref.~\cite{szolnoki2016competition}. The shaded region indicates the range of network sizes where mutation results align with those in large populations, effectively resolving finite-size effects. Mutation rate is fixed at $\mu=10^{-6}$. The parameters are set as $r=2.2$ and $\gamma=0.1$ for panel (a) and (c),  $r=2.2$, and $\gamma=0.1$ for panel (b) and (d).}
    \label{figm1_n}
\end{figure*}

\end{document}